# AN EMPIRICAL EVALUATION FOR THE INTRUSION DETECTION FEATURES BASED ON MACHINE LEARNING AND FEATURE SELECTION METHODS

**MOUHAMMD ALKASASSBEH**

Computer Science Department, Information Technology College, Mutah University, Jordan

E-mail: mouhammd.alkasassbeh@mutah.edu.jo

**ABSTRACT**

Despite the great developments in information technology, particularly the Internet, computer networks, global information exchange, and its positive impact in all areas of daily life, it has also contributed to the development of penetration and intrusion which forms a high risk to the security of information organizations, government agencies, and causes large economic losses. There are many techniques designed for protection such as firewall and intrusion detection systems (IDS). IDS is a set of software and/or hardware techniques used to detect hacker's activities in computer systems. Two types of anomalies are used in IDS to detect intrusive activities different from normal user behavior. Misuse relies on the knowledge base that contains all known attack techniques and intrusion is discovered through research in this knowledge base. Artificial intelligence techniques have been introduced to improve the performance of these systems. The importance of IDS is to identify unauthorized access attempting to compromise confidentiality, integrity or availability of the computer network. This paper investigates the Intrusion Detection (ID) problem using three machine learning algorithms namely, BayesNet algorithm, Multi-Layer Perceptron (MLP), and Support Vector Machine (SVM). The algorithms are applied on a real, Management Information Based (MIB) dataset that is collected from real life environment. To enhance the detection process accuracy, a set of feature selection approaches is used; Infogain (IG), ReleifF (RF), and Genetic Search (GS). Our experiments show that the three feature selection methods have enhanced the classification performance. GS with bayesNet, MLP and SVM give high accuracy rates, more specifically the BayesNet with the GS accuracy rate is 99.9%.

## 1. INTRODUCTION

Lately, there has been a continuous increase in security threats which pose a serious challenge to the Internet and prosperity of the information technology (IT) business section. Denial of Service (DoS) attack is considered one of the strongest and most effective attacks. Since the summer of 1999, many DoS attacks were carried out against different organizations world-wide. These attacks have crippled the economy and have even caused some businesses to leave the market. Since September 2012, online banking websites of nine major U.S. banks have experienced a series of such attacks [1], leading to financial losses and making it difficult and costly to restore services and revert to normal functioning. Therefore, it is vital to study such attacks and deal with them to defend businesses. Intrusion detection system (IDS) is a process of monitoring the network traffic from any occurring abnormal action or behavior and comparing it with normal activities to identify signs of intrusion. Intrusion refers to a malicious activity aimed at compromising the confidentiality, integrity, and availability of network components in an attempt to disrupt the security policy of networks [2]. Misuse Intrusion Detection (MID) and Anomaly Intrusion Detection (AID) are the two main detection methods in IDS. MID is also known as signature-based detection. It works on pattern matching for known intrusion patterns to detect attacks but is likely to miss and not detect new types of attacks. In AID, anomaly detection is based on identifying abnormal behavior of hosts or networks. It needs the previous construction of normal behaviors which are collected over a period time under normal operations. The advantage with AID systems is the ability to detect unusual attacks without specific previous knowledge about them. Currently, researchers have mainly focused on AID because it can detect known and unknown attacks. Figure 1 shows an overview of IDS systems. This study is, specifically, focused on network anomaly detection methods.

Furthermore, network anomaly is a deviation in the normal behavior of a network in the presence of any intruder in a network or due to network overload.

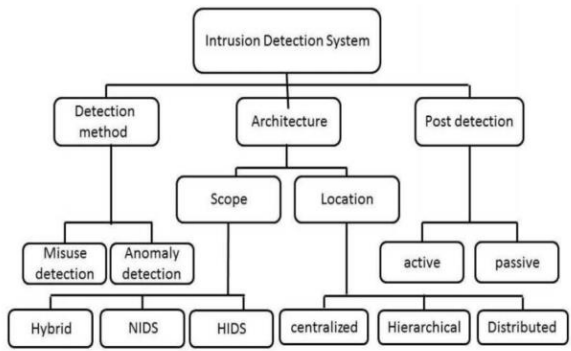

*Figure 1: IDS systems [3]*

These anomalous events disrupt the normal functionality of network services. Normal network behavior can be characterized by various factors, such as the type of network data to be measured, traffic volume of the network, and the type of applications that are running on the network [2]. To detect network traffic anomalies, traffic features describing network traffic behavior is used to create a normal network traffic model. Model generation algorithms or mathematical models can generate this model from the training data [4]. A network attack is any process or method that is used to carry out malicious attempts against computer networks to compromise network security [5]. Attackers consider various distinct stages when they want to carry out attacks, starting from the initial motivation of the attackers to the final execution of the attacks [6]. A DOS attack is the most attractive type of attack to attackers. The focus of this study is on the detection of attacks, specifically DOS flooding attacks and brute force attack. The most common types of DOS Flooding Attack are Transmission Control Protocol-Synchronize (TCP-SYN) Attack, User Datagram Protocol (UDP) Flood Attack, Internet Control Message Protocol-Echo (ICMP-Echo) Attack, HyperText Transfer Protocol (HTTP) Flood Attack, Slowloris Attack, and Slowpost Attack. On the other hand, in brute force (password cracking) attack, attackers attempt to log in to a user account by continuously trying different passwords on the victim until they find the correct password. For example, a recent article [7] in 2014 reported that a large number of user passwords were very weak, such as "123456", "sunshine", "qwerty", etc. The correct password is easily divulged through brute force attack. All the details of the attacks and the dataset used in this study was taken from previous research conducted by me and other researchers [8]. There are various approaches detect network anomalies [2] [9], for example, statistical approach, rule-modelling approach, threshold approach, and machine learning (ML) approach. In statistical approach, anomalies are detected from statistics; depending on the analysis of data, a decision is made whether this is normal or abnormal behavior. In the rule-modelling approach, anomalous activities are detected when system-defined rules are broken. In the threshold approach, abnormal activities are detected when thresholds defined for data deviation monitoring are crossed. In the last approach, ML, a model is constructed based on past data or behaviors. Normal and abnormal data are collected from a network under normal and abnormal conditions. Data is then labeled for training to train the model. The biggest advantage of the ML method is that it can be adopted for unknown attacks and can be adapted to any changes in the network traffic. ML is the approach used most for abnormal detection [10].

The contributions in this paper include:

(1) Investigating the Intrusion Detection (ID) problem using three machine learning algorithms namely, BayesNet algorithm, Multi-Layer Perceptron (MLP), and Support Vector Machine (SVM).

(2) Applying these algorithms on a real, Management Information Based (MIB) dataset that is collected from real life environment.

(3) Enhancing the detection process accuracy by using a set of feature selection approaches such as, Infogain (IG), ReleifF (RF), and Genetic Search (GS). (4) The experiments show that the three feature selection methods have enhanced the classification performance, specifically the accuracy rate of the BayesNet with the GS is 99.9%.

This paper is organized as follows. The introduction is presented in Section 1. Section 2 provides a literature review covering various techniques used to solve the IDS problem. Section 3 briefly explains the nature of the data set used in the experiments. The three ML algorithms namely, BayesNet, Multi-Layer Perceptron (MLP), and sector vector machine (SVM) are defined in Section 4. Feature selection methods based on Ranker and genetic search (GS) methods are presented in Section 5. Methods of evaluating the developed model are presented in Section 6. Experimental results are presented in Section 7. Finally, the conclusions are presented in Section 8.

## 2. RELATED WORK

Over the past decade, many researchers extensively studied the detection of anomalies and attacks on computer networks [11]. The vast majority of the solutions presented to date within the scope of network anomalies and attack detection have focused on analyzing raw traffic features, such as interarrival time, number of packets, Internet Protocol (IP) addresses, ports, network flow, etc. Other solutions based on Simple Network Management Protocol Management Information Base (SNMP MIB) as a data source were proposed to detect network anomalies.

In this section, previous related works on anomaly detection using SNMP-MIB will be reviewed. An earlier attempt to exploit SNMP for network security was stated in [12], authors proposed a methodology for the early detection of Distributed Denial of Service (DDoS) attacks by applying statistical tests for causality to extract MIB variables that contain signs of attacks. Their method relies on 91 MIB variables from five groupsIP, ICMP, TCP, UDP, and SNMPcollected periodically from the target and the attacker participating in attacks. Three types of DDoS attacks were conducted. Their work has shown that it is possible to extract a precursor to a DDoS attack using MIB traffic variables and to detect these attacks before the target is shut down with about 1% rate of false alarms.

In another work [13], presented an experimental analysis of a proactive detection method based on statistical tests for causality to detect early DDoS attacks. They implemented a system based on SNMP MIB variables which collected four MIB variables corresponding to four groupsIP, ICMP, TCP, and UDP at 5 second sampling intervals and tested five DDoS attacks. The results of their experiments achieved a high rate of success in detecting DDoS attacks.

In [14], the authors applied an Auto Regressive (AR) method to a time series obtained from five MIB variables corresponding to IP and IF groups, and conducted a sequential hypothesis test to detect network anomalies. They evaluated their detection method by conducting real experiments involving SYN flood, ICMP flood, and Smurf and UDP flood attacks. They claimed their method satisfactorily detected traffic anomalies. Also in [5], they developed an anomaly detection system based on network traffic collected periodically from SNMP MIB data which included four MIB variables from an interface group IF. They performed two DOS attacks, Smurf and SYN-Flood attacks, and a scan attack to evaluate the developed system in a real experiment. They discovered this system could efficiently detect flooding attacks. In [6], the authors proposed a fast and lightweight algorithm for intrusion detection, especially for traffic flooding attacks detection, based on the correlation of SNMP MIB variables. They selected 16 MIB-II variables from 6 groups to be used for attacks detection. TCP-SYN, UDP, and ICMP flooding attacks were used to test the performance of the proposed detection algorithm in a real experiment. They proved their proposed work could accurately detect all types of attacks with extremely low false positive (FP) and false negative (FN) rates.

In [15] and [16], they utilized a ML approach for network intrusion detection based on SNMP MIB data. Both works proposed a fast and lightweight system that can detect and classify traffic flooding attacks based on SVM. They gathered 13 SNMP MIB variables corresponding to 4 MIB groupsIP, ICMP, TCP, and UDPat 15-second sampling intervals from SNMP agents in real experiments. The proposed system was constructed in a hierarchical SVM-based structure for attack traffic detection and classification into different types of attacks: TCP-SYN, UDP, and ICMP floods. The proposed system effectively detect known and new attacks. They also concluded they had achieved fast detection with a high detection accuracy of 99.27% and with a low rate of false alarms using SVM and key MIB variables selected from a correlation feature selection (CFS) mechanism.

In [17], authors presented an anomaly detection system using a decision tree algorithm with SNMP MIB data to detect an early Trojan attack. They focused mainly on Host Resources MIB variables related to services and running software, which can also be an indicator of changes in system performance.

In [18], the researchers presented an intrusion detection system based on SNMP MIB and ML approaches. Their system consisted of three modules. The first module was for selecting key MIB variables from three classification algorithms: C4.5, RIPPER and attribute selection. The second module was for generating an intrusion detection model based on the chosen variables and detecting DOS/DDOS attacks in real time in the third module. Four MIB groupsIP, ICMP, TCP, and UDPhad been used involving TCP-SYN, UDP, and ICMP flood attacks. The tested system achieved a 99.03% accuracy rate by using a neural network algorithm among three classification algorithms: neural network, Bayesian network, and C4.5.

In [19], the authors proposed a system namely, the Protocol Independent Detection and Classification (PIDC), to detect and classify Distributed Reflection Denial of Service (DRDOS) attacks, such as a domain name system (DNS) attack and a TCP

SYN reflection flooding attack. They captured 13 MIB variables from the TCP and DNS groups, using a rank correlation-based detection algorithm to determine the relationship between these variables. The C4.5 classification algorithm was then used with MIB variables to classify the type of attacks that were considered in this research. Their method achieved a 99% true positive rate and a 1% false positive rate in detecting reflected attacks.

In [20], they authors focused on a statistical signal processing technique to solve the problem of network anomaly detection using three MIB variables from the IP group. This technique is based on abrupt change detection in each of the MIB variables by using a hypothesis test based on the Generalized Likelihood Ratio (GLR). Any deviations in the behavior of MIB variables are then correlated according to their characteristics. Based on the case studies presented, Protocol Implementation Errors, File Server Failure, Network Access Problems, and Runaway Processes, they concluded that applying signal processing techniques could add significant advantages in solving the problem of network anomaly detection.

In [21] and [22], Mobile Agent (MA) has been used to read the MIB variables from the local machines to overcome the limitations of centralized IDSs. This work showed that the statistical methods based on Wiener filter and the MA technology can be combined to detect network intrusion. The Wiener statistical filtering takes advantage of the correlation matrix between the input MIB variables and the crosscorrelation with the desired MIB variables to detect abnormal situations in the traffic. The new model managed to detect all the attacks in light network traffic but in heavy traffic it had some difficulty in detecting attacks. The chosen MIB was for IF and IP groups and the scenarios were decoy port-scan, buffer overflow, brute force attack and null session attack.

In [23], they developed a statistical approach based on information entropy for network anomaly detection using real SNMP MIB data. Their work considered four case studies of network anomalies to evaluate their approach using only a set of MIB variables corresponding to the IF group: 100% cup occupancy, database shutdown failure, application server shutdown failure, and database failure.

As shown above, the SNMP protocol and MIB variables play a good role in detecting network attacks, however, the MIB database has 171 variables. These variables fall into 10 groups depending on the type of node and the management functions. In the case of a server, a switch or a router, there are different groups of MIB variables of each to observe statistical and status information. This raises the following questions, which will be answered in this study: Which one of these variables is important and which one is not? Which variable has a high impact in detecting the attacks? What is the most important group of MIB that gives a good and clear sign of attacks as there is no single MIB variable responsible to give an indication of any attack?

3. DATA SET

SNMP-MIB data is rich data source that provides clear statistical information about the current network device status. As the SNMP-MIB is a widely deployed protocol in most network devices, it does not need any new hardware or any effort for configuration. By reading MIB data some of the problems and challenges in intrusion detection can be avoided. The data used in this study is described in [8]. It is a new dataset with up-to-date attacks. The dataset contains 4,998 connection records. The data distributed in eight main classes are described in Table 1.

*Table 1: Traffic types and number of generated records*

| Type of traffic | Number of records |
|---|---|
| Normal | 600 |
| TCP-SYN | 960 |
| UDP flood | 773 |
| ICMP-ECHO | 632 |
| HTTP flood | 573 |
| Slowloris | 780 |
| Slowpost | 480 |
| Brute Force | 200 |
| Total | 4998 |

The MIB dataset was collected from a router with 34 MIB variables from 5 MIB groups in MIB-II. The groups are IF, IP, TCP, UDP and ICMP. The groups and their variables are listed in Table 2.

*Table 2: MIB variables in each group*

| Group name | Variable number | Variable name |
|---|---|---|
| IF | 1 | ifInOctets |
| | 2 | ifOutOctets |
| | 3 | ifoutDiscards |
| | 4 | ifInUcastPkts |
| | 5 | ifInNUcastPkts |
| | 6 | ifInDiscards |
| | 7 | ifOutUcastPkts |
| | 8 | ifOutNUcastPkts |
| TCP | 1 | ifInOctets |
| | 9 | tcpOutRsts |
| | 10 | tcpInSegs |
| | 11 | tcpOutSegs |
| | 12 | tcpPassiveOpens |
| | 13 | tcpRetransSegs |
| | 14 | tcpCurrEstab |
| | 15 | tcpEstabResets |
| | 16 | tcpActiveOpens |
| UDP | 17 | udpInDatagrams |
| | 18 | udpOutDatagrams |
| | 19 | udpInErrors |
| | 20 | udpNoPorts |
| IP | 21 | ipInReceives |
| | 22 | ipInDelivers |
| | 23 | ipOutRequests |
| | 24 | ipOutDiscards |
| | 25 | ipInDiscards |
| | 26 | ipForwDatagrams |
| | 27 | ipOutNoRoutes |
| | 28 | ipInAddrErrors |
| ICMP | 21 | ipInReceives |
| | 29 | icmpInMsgs |
| | 30 | icmpInDestUnreachs |
| | 31 | icmpOutMsgs |
| | 32 | icmpOutDestUnreachs |
| | 33 | icmpInEchos |
| | 34 | icmpOutEchoReps |

## 4. MACHINE LEARNING TECHNIQUES

In this section, we introduce the ML techniques used to classify the DoS attacks.

### 4.1 BayesNet

A Bayesian network is a Directed Acyclic Graph (DAG) known as BayesNet in Waikato Environment for Knowledge Analysis (WEKA) tool [24]. A Bayesian network is enormously beneficial as it helps to understand the world researchers model. It may be the best to use in various areas of life when modelling a mysterious fact and in the state of decision nets or wherever it is good to make intelligent, justifiable, and quantifiable decisions that will enhance performance of classification. Briefly, BayesNet is helpful for diagnosis, prediction, modeling, monitoring, and classification [25].

The main idea of the Bayesian classifier consists of two phases: in the first, if an agent has an idea and knows the class it can predict the values of the other features; in the second, if the agent does not have an idea or does not know the class the Bayes rule is used to predict the given class. In a Bayesian classifier, the learning agent builds a probabilistic model of the features and uses this model to predict the classes. The Bayesian Network has been used as a classifier for the following reasons:

- Probabilistic learning which calculates the clear probabilities for assumption.
- Incremental which is a prior knowledge and possible to add to data viewing.
- Probabilistic prediction, which can predict more than one hypothesis, weighted by the probabilities.

The theory of the Bayesian Network is shown in Equation 1, where the symbol $D$ indicates the training data, a posteriori probability of hypothesis h [26].

$$p(h \mid D) = \frac{p(D \mid h)p(h)}{p(D)} \quad (1)$$

Where $P(h/D)$ refers to posterior probability, $P(D/h)$ refers to condition probability, $P(h)$ is prior probability of $h$ and $P(D)$ refers to marginal probability of $D$.

### 4.2 Multi-Layer Perceptron

Artificial Neural Network (ANN) is one of the most popular machine learning algorithms. It was inspired by the neural structure of the human brain which contains billions of neurons connected to each input by synaptic weights and the output is generated on the other side [27]). This concept is applied in ANN in computer systems to solve classification and regression problems. MLP is the most common model and widely used class of ANN. It consists of one input layer, one output layer, and one or more hidden layers. MLP aims to create a relationship that maps a set of inputs into a set of suitable outputs, then the MLP model can be used to extract unknown outputs [27].

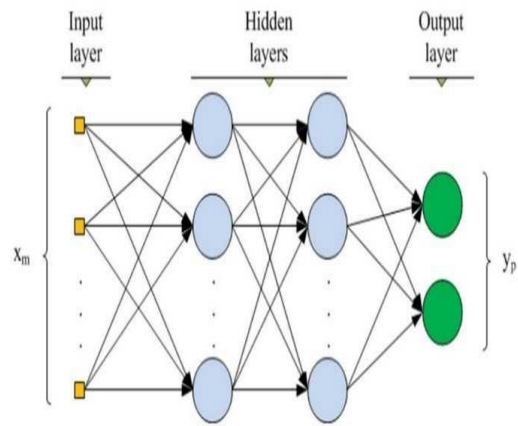

*Figure 2: Simple MLP architecture with three layers*

Figure. 2 shows a simple MLP neural network with two hidden layers. Where a bias parameter with initial weight value is added to the input layer. The input values are entered into the input layer, and then multiplied by the weight values as they are passed from the input layer to the first hidden layer. At the hidden layer in each neuron the multiplications are added then processed by a transfer function. The same process is applied to the output of each hidden layer until the last hidden layer. The last hidden layers outputs are multiplied by the weights then processed by transfer function within the output layer to produce the output of neural network [27].

Different transfer functions are used in neural networks. In this paper, the most common transfer function, the logistic sigmoid function, was used. The function is defined as:

$$F(x) = \frac{1}{1 + e^{(-\beta x)}} \quad (2)$$

Many algorithms can be used in the MLP during the training phase. The Back Propagate algorithm (BP) is one of the most common neural network algorithms. In the BP algorithm input values are entered repeatedly into the networks where each output value is computed to obtain the preferred output and an error is computed. The error is then back propagated to the network to update the weights to reduce error with each replication and the desired output moves closer to obtainment [28].

### 4.3 Support Vector Machine

Support Vector Machine (SVM) is an advanced ML method where it overtakes many other classic ML techniques in various fields. SVM is an incredibly efficient method for classification; it determines the most optimal separating hyper-plane among classes depending on the training data set. To demonstrate this technique, suppose there are two linearly separable classes in a certain d-dimensional space with the use of training vectors related to two classes, $\{x_i, y_i\}$, in which $x_i \in R^d$ signifies vectors in the d-dimensional space, while $y_i \in \{-1,+1\}$ represents a class label. The purpose in the design of a hyper-plane is to simplify data in an accurate way where this hyperplane is the one that leaves the extreme margin from both classes [29]. The main idea of the SVM technique is to find the hyper-plane which has the most extreme margin towards the sample object, where the margin value and the probability to inaccurately classify a feature vector are inversely related to each other. The following equation can be used to define a hyper-plane [29]:

$$(w.x) + b = U \quad (3)$$

where $w$ is a normal to the hyper-plane, $x$ represents a feature vector that lies on that hyper-plane, and $b$ represents the bias in which $\frac{|b|}{\|w\|}$ represents a perpendicular distance among the origin and the middle point of the hyper-plane (Figure 4). The purpose is to separate among two classes: open circle that stands for the class label -1 and the solid circle which stands for the +1. The lying circles on both planes, P1 and P2, represent the support vectors in which the optimal hyper-plane located among those two plans are parallel to each other. The margin among those planes is $\frac{|2|}{\|w\|}$. The SVM technique should maximize the hyper-plane margin to get enhanced generalization. The following formulas can be used to describe the hyper-plane of the two classes [31]:

$$(w.x) + b = +1 \text{ for class} = +1 \quad (4)$$
$$(w.x) + b = -1 \text{ for class} = -1 \quad (5)$$

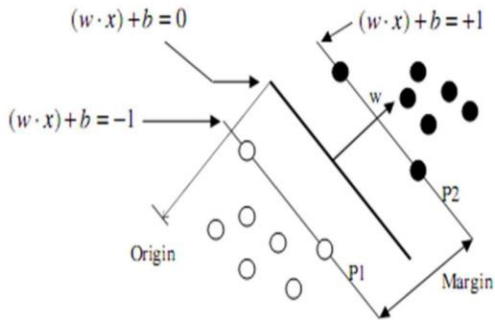

*Figure 3: SVM classification basics [30]*

Practically, classes are not linearly separated. Thus, the input space must be mapped into another feature space with high dimensionality. More clearly, input vectors, such as low-level feature vectors, are mapped into a feature space H using a nonlinear conversion: $\varphi : R^d \to H$ Thus, the optimal hyper-plane is generated in that high dimensional feature space with the use of kernel function, $K(x_i, x_j)$ that is generated among two input vectors: $x_i$ and $x_j$. This formula can be written as follows [32]:

$$K(x_i, x_j) = \varphi(x_i).\varphi(x_j) \quad (6)$$

Polynomial kernel is one of the most common mappings. This formula is described as:

$$K(x_i, x_j) = (x_i.x_j + 1)^d \quad (7)$$

where $d$ represents the polynomial degree. Another common mapping is the Radial Basis Function (RBF) kernel described as [32]:

$$K(x_i, x_j) = e^{\frac{\|x_i - x_j\|^2}{2\sigma}} \quad (8)$$

where, $\sigma$ stands for the Gaussian sigma. As described above, the SVM technique is developed to solve binary classification problems with two class labels only: +1 and 1. This technique can be enhanced more for use with multi-class problems. One against all and one against one are the two approaches for SVM multi-class classification. The one against all approach includes the construction of SVMs among each class and other classes.

### 5. FEATURE SELECTION

The main objective of the feature selection is to reduce the computation time and improve the model performance by reducing the size of the data the model needs for learning and for the setup stage. In a simple way, the feature selection retains the most important fields and removes irrelevant ones while it still provides a good end result [33].

## 5.1 Feature selection methods

A summary of feature selection techniques is presented in Figure. 4. There are three main categories of feature selection methods based on the evaluation criteria: filter, wrapper, and hybrid methods. The filter method depends on the general characteristics of the training data set to select features with independence of any predictor. In the wrapper method, the selection process involves optimized classifiers to provide the most important subsets. Although better results occur with this method because classifiers are used as part of the selection process computational complexity is higher. Contrariwise , filter method is sufficient when dealing with large data sets over the wrapper method. The advantage of the hybrid method the combination of filter and wrapper is their application during the selection process at different stages [34]. The aim of this work is to evalute the various existing attribute selection methods in terms of detection rate (accuracy) and computational performance. Out of the total 34 features used in detecting intrusion some of these features will be optional in detecting intrusions. Consequently, the main features are extracted from the main set that are especially effective in detecting intrusions which will be discussed in the results section. Herein, both methods were used to make a comparison between them and the filter and the wrapper methods were chosen. InfoGain and and ReleifF were chosen for the filter method and GeneticSearch was chosen for the wrapper method.
- Attribute evaluators are used to rank all the features in the main set giving to some metrics.

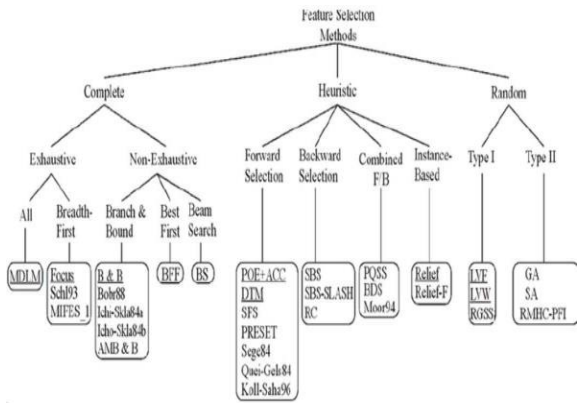

*Figure 4: Summary of feature selection methods [34]*

There are numerous attribute evaluators available in the WEKA tool.
- InfoGain: It expresses the importance of a feature or attribute and determines which attribute is the most useful for discriminating between the classes to be learned. This piece of information is also calculated on training data set. Moreover, it can help in choosing the best split; if it has a high value then this split is good, otherwise the split is not good enough. It can be calculated by the data achieved from entropy:

- *InfoGain=entropy(parent)−[averageentropy( children)].*

- ReleifF: It was first described by Robnik and Rendell [35] as a simple and effective approach for attribute ranking. The output of the algorithm is a value between - 1 and 1 for each feature with more positive numbers indicating more important or weighted attributes. The weight of an attribute is updated iteratively. A sample is selected from the data set, and the nearest neighboring sample that belongs to the same class and the nearest neighboring sample that belongs to the other class is recognized. Any change in any attribute value leads to weighting of the attribute based on that change which could be responsible for the class change. Instead a change in attribute value with no change in class leads to down weighting of the attribute. This process is performed for either a random set of samples or for the whole data set. The attribute weight estimated by ReliefF has a probabilistic explanation. It is proportional to the difference between two conditional probabilities, namely, the probability of the attribute's value being different conditioned on the given nearest miss and nearest hit respectively [36].

- Search Methods
  - Ranker The Ranker search method ranks features by their individual evaluations. It sorts them from the highest to the lowest based on their importance.
  - Genetic Algorithm (GA) is an adaptive heuristic search algorithm based on the evolutionary concepts of natural selection

[4]. This concept is used to solve optimization problems in intelligent exploitation. GA exploits historical information in the domain to direct the search into a region of improved performance within the search space. GA uses in its mechanism an individual called chromosome. These chromosomes are created randomly in the initial population. Each one denotes one possible solution to the problem [37]. The newly produced solutions grow over time to give an optimal solution. In case of feature selection problems, the produced solution or outcome is a binary string that demonstrates the selected features. Each feature in the produced solution subset represents its importance [4]. The first generation is selected randomly from the features set. For following generations, they are produced using genetic operators like crossover (single point or multipoint) and mutation. The new generated subset of the features is evaluated using fitness function. The superior subsets have a stronger chance to be selected for the new generation. Algorithm 1 shows, generally, how GS approach works.

Algorithm 1: Genetic search algorithm [4].

| | |
|---|---|
| 1 | Begin by randomly generating a two initial population $P$. |
| 2 | Calculate $e(x)$ for each member $x \in P$. |
| 3 | Define a probability distribution $p$ over the members of $P$ where $p(x)e(x)$. |
| 4 | Select two population members x and y with respect to $P$. |
| 5 | Apply crossover to $x$ and $y$ to produce new population members $\hat{x}$ and $\hat{y}$. |
| 6 | Apply mutation to $\hat{x}$ and $\hat{y}$. |
| 7 | Insert $\hat{x}$ and $\hat{y}$ into $\hat{p}$ (the next generation). |
| 8 | If $|\hat{p}| < |p|$, goto 4. |
| 9 | Let $p \leftarrow \hat{p}$. |
| 10 | If there are more generations, goto 2. |
| 11 | Return $x \in p$ for which $e(x)$ is highest. |

## 6. MODEL EVALUATION

A set of performance evaluation functions was explored to check the developed models performance:

- Confusion Matrix: Table 3 provides a visualization of the performance of the classification model. It contains information about actual and predicted classifications of the model.

*Table 3: Confusion Matrix for two classes*

| | Predicted Class | | |
|---|---|---|---|
| | | Positive | Negative |
| Actual Class | Positive | TP | FP |
| | Negative | TN | FN |

A confusion matrix has four measurement factors that are commonly used to evaluate the performance of a classification model:

- True Positive (TP) refers to the correct prediction rate of the positive samples.
- False Positive (FP) refers to the ratio of negative samples incorrectly classified as positive.
- True Negative (TN) refers to the ratio of negative samples correctly classified as negative.
- False Negative (FN) refers to the ratio of positive samples incorrectly classified as negative.

- Accuracy: Accuracy refers to the percentage of correctly classified instances. It is the most basic metrics for performance and it determines how accurately the classification model is evaluating all classes. Accuracy is calculated by using a confusion matrix factors with the following equation:

$$Accuracy = \frac{(TP + FN)}{(TP + FP + TN + FN)} \quad (9)$$

- Precision: Precision is the ratio of the predicted positive samples correctly classified. Precision is calculated using the following equation:

$$Precision = \frac{(TP)}{(TP + FP)} \quad (10)$$

- Recall: Recall refers to the true positive rate. It is the ratio of positive samples classified correctly as positive. Recall is calculated using the following equation:

$$Recall = \frac{(TP)}{(TP + FN)} \quad (11)$$

- F-Measure: F-measure is a measure of a classification models accuracy depending on the precision and the recall metrics. It is a weighted average of both Precision and Recall metrics. F-Measure is calculated using the following equation:

$$F - measure = 2 * \frac{(Precision * Recall)}{(Precision + Recall)} \quad (12)$$

## 7. EXPERIMENTAL RESULTS

In this work, the dataset described in section 3 was used. Table 1 showed the type of data that was used and the number of samples for each class (attack type). The dataset has eight different attack types which reflect most of the current attack scenarios. Weka 3.6 has been used as a comparative analysis. Different combinations of feature selection have been used in this study: Ranker+InfoGain, Ranker+ReleifF, and Wrapper+GeneticSearch. The details of these combinations and their selected features subsets are described in Table 4.

*Table 4: Feature subsets*

| Method | Features |
|---|---|
| InfoGain | 8, 24, 11, 10, 12, 7, 31, 29, 4, 26, 21, 30, 33,34, 5, 2, 20, 32, 25,28, 22, 17, 23, 18, 19, 6,3, 15, 13, 16, 9, 27, 14, 1 |
| ReliefF | 27,32,28,9,5,8,20,25,4,15,26,21,18,23, 2, 19, 7, 17, 11, 22, 12,10, 31, 34, 33,29, 30, 13, 24, 16, 6, 3, 1, 14 |
| Genetic Search-SVM | 3, 4, 5, 6, 7, 8, 9, 10, 11, 12, 14, 15, 16, 17,18, 22, 23, 24, 25, 26, 27, 28, 29, 30, 31,32, 33, 34 |
| Genetic Search-MLP | 1, 3, 5, 7, 9, 10, 13, 15, 16, 18, 19, 20, 21,23, 24, 25, 26, 28, 29, 32 |
| Genetic Search-BayesNet | 3, 6, 7, 8, 9, 17, 19, 24, 26, 30, 34 |

For all experiments, the models are tested using a 10-fold cross validation because it decreases the variance of the estimation. In this technique, the training dataset is splited into 10 subsets each subset is then tested on the model that trained on the other nine subsets. Although the process is repeated 10 times each test subset is only used once.

Table 6 shows the performance of the three main classifiers (BaysNet, MLP, and SVM) in terms of Recall, Precision, F-measure and accuracy rate. The first part of the table is the BayesNet results. The first raw is the results of all features (34). The algorithm achieved very high results with a 99.5% accuracy rate taking 0.14 seconds to build the model. Using the same algorithm with feature selection method the GS produced better results with just 11 features, the accuracy produced an exceedingly high result of 99.9% in 0.05 seconds. For the MLP, the results for all of the features was 99.40% in 32.63 seconds.

Although it was a good result it took a relatively long time to build the MLP model. As the results show, GS again produces better results than IG and RF. MLP with GR method gives an accuracy rate of 99.60% in 14.42 seconds with 20 features. SVM with GS method has similar results with a 99.1% accuracy rate with 28 features and taking 0.75 seconds to build the model. Over the three algorithms, BayesNet with GS has an exceptionally high accuracy within an incredibly short time frame with limited numbers of features.

The results in Table 5 and Figure 5 provide the accuracy for each MIB group. In this study, GS was again used with the same algorithms to discover the most important group in MIB. The results show that the most significant group is the IP group followed by the IF group with an accuracy of 99.80% and 99.40% respectively both using the BayesNet algorithm. The study analysis was extended and the most repeated features in all search methods were collected and fed into the tested algorithms. There were only five features: 3, 7, 9, 24, and 26. As expected with BayesNet, the results are incredibly high with an accuracy of 99.8% (Table 7 ).

In the last case of this study, a MIB data was used based on this researchers previous work. Detailed explanation on the selection of MIB variables is available in [21] [22]. Again, the BayesNet has the highest accuracy with 98.70% accuracy using the chosen MIBs: 5, 7, 8, 21, 22, and 23 (Table 8).

*Table 5: Models evaluation results*

| ALGORITHM | ACC | Precision | Recall | F-Measure | Time Taken (seconds) |
|---|---|---|---|---|---|
| BayesNet | 99.50% | 99.50% | 99.50% | 99.50% | 0.14 |
| BayesNet+infogain+top5 | 98.40% | 98.40% | 98.40% | 98.40% | 0.02 |
| BayesNet+reliefF+top5 | 83.40% | 87.30% | 83.40% | 83.80% | 0.02 |
| BayesNet+infogain+top10 | 99.10% | 99.20% | 99.10% | 99.10% | 0.06 |
| BayesNet+reliefF+top10 | 99.20% | 99.30% | 99.20% | 99.20% | 0.14 |
| BayesNet+geneticsearch | 99.90% | 99.90% | 99.90% | 99.90% | 0.05 |
| MLP | 99.40% | 99.40% | 99.40% | 99.40% | 32.63 |
| MLP+infogain+top5 | 88.90% | 89.70% | 88.90% | 88.80% | 5.78 |
| MLP+reliefF+top5 | 96.50% | 96.80% | 96.50% | 96.40% | 5.85 |
| MLP+infogain+top10 | 93.90% | 94.30% | 93.90% | 93.70% | 8.99 |
| MLP+reliefF+top10 | 98.40% | 98.40% | 98.40% | 98.40% | 9.48 |
| MLP+geneticsearch | 99.60% | 99.60% | 99.60% | 99.60% | 14.42 |
| SVM | 97.90% | 97.90% | 97.90% | 97.80% | 0.67 |
| SVM+infogain+top5 | 63.20% | 73.50% | 63.20% | 63.70% | 0.62 |
| SVM+reliefF+top5 | 66.60% | 61.70% | 66.60% | 62.70% | 0.42 |
| SVM+infogain+top10 | 84.20% | 87.90% | 84.20% | 84.10% | 0.53 |
| SVM+reliefF+top10 | 88.90% | 92.00% | 88.90% | 89.10% | 0.8 |
| SVM+geneticsearch | 99.10% | 99.10% | 99.10% | 99.10% | 0.75 |

*Table 6: Results based on each MIB group*

| Group | Algorithm | ACC | Time Taken (seconds) |
|---|---|---|---|
| IF Group | BayesNet+geneticsearch | 99.40% | 0.06 |
| | MLP+geneticsearch | 97.30% | 13.8 |
| | SVM+geneticsearch | 71.60% | 0.9 |
| TCP Group | BayesNet+geneticsearch | 83.90% | 0.02 |
| | MLP+geneticsearch | 83.60% | 7.88 |
| | SVM+geneticsearch | 69.90% | 0.64 |

| Group | Algorithm | ACC | Time |
|---|---|---|---|
| UDP Group | BayesNet+geneticsearch | 77.20% | 0.02 |
| | MLP+geneticsearch | 94.90% | 7.65 |
| | SVM+geneticsearch | 65.90% | 0.6 |
| IP Group | BayesNet+geneticsearch | 99.80% | 0.07 |
| | MLP+geneticsearch | 98.10% | 6.99 |
| | SVM+geneticsearch | 85.90% | 0.67 |
| ICMP Group | BayesNet+geneticsearch | 87.00% | 0.02 |
| | MLP+geneticsearch | 80.80% | 7.47 |
| | SVM+geneticsearch | 62.30% | 1.22 |

*Table 7: The most frequent features in all subsets*

| Algorithm | ACC | Time Taken (seconds) |
|---|---|---|
| BayesNet | 99.80% | 0.03 |
| MLP | 94.30% | 6.58 |
| SVM | 72.90% | 0.75 |

## 8. CONCLUSIONS

In this paper, three models were designed to solve the intrusion detection problem using BayesNet, MLP and SVM. The number of attacks was classified using above methods. To enhance the performance of the proposed models and to speed up the detection process, a set of features was selected using Ranker and GS methods. A comparison between the models before and after feature selection was provided. Our findings show that the models were capable of reducing the complexity while retaining acceptable detection accuracy. The BayesNet algorithm with GS method achieved the highest classification accuracy compared to other search techniques. It should be noted that, neither a single MIB variable is responsible to catch different type of attacks nor a specific MIB group such as, IF or IP group. Hence, our work based on the selected dataset showed that the variables namely, ifoutDiscards, ifOutUcastPkts, tcpOutRsts, ipOutDiscards, ipForwDatagrams, play a viatale role in catuching all examined attackes with high accuracy rate.

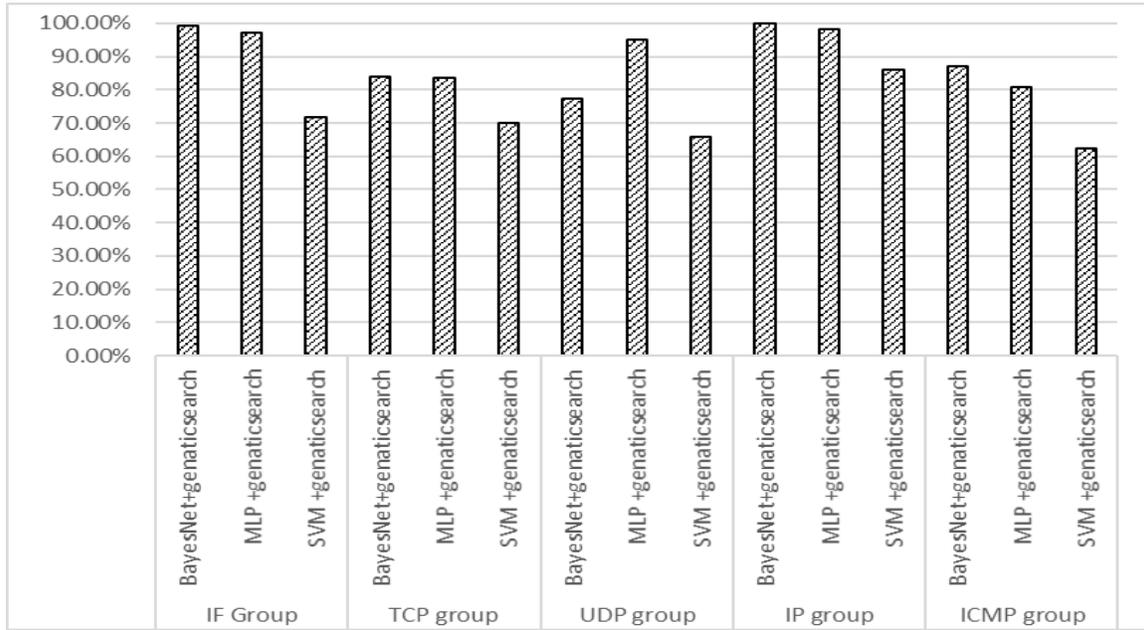

*Figure 5: Accuracy based on each group*

*Table 8: Results based on this researchers previous work [21] [22]*

| Algorithm | ACC | Time Taken (seconds) |
|---|---|---|
| BayesNet | 98.70% | 0.15 |
| MLP | 90.80% | 7.4 |
| SVM | 60.50% | 0.5 |